\documentclass[trackchanges, twocolumn]{aastex7}
\usepackage[dvipsnames]{xcolor}
\usepackage{subfiles}
\usepackage{rotating}
\usepackage{url}
\usepackage{amsmath}
\usepackage{multirow}
\usepackage{booktabs}
\usepackage{tabularx}
\usepackage{graphicx}
\usepackage{import} 

\colorlet{KMcolor}{WildStrawberry}

\begin{document}

\title{Considering Numerical Relativity Surrogate Models for AGN Channel Merger Dynamics}

\author[0009-0005-5038-3171]{Shawn Ray}
\affiliation{Department of Astrophysics, American Museum of Natural History, New York, NY 10024, USA}
\affiliation{Graduate Center, City University of New York, 365 5th Avenue, New York, NY 10016, USA}
\email{shawn.ray81@gc.cuny.edu}

\author[0000-0003-0276-3856]{Keefe Mitman}
\affiliation{Cornell Center for Astrophysics and Planetary Science, Cornell University, Ithaca, NY 14853, USA}
\email{kem343@cornell.edu}

\author[0000-0001-7099-765X]{Vera Delfavero}
\affiliation{Canadian Institute for Theoretical Astrophysics, University of Toronto, 60 St George St, Toronto, ON M5S 3H8, Canada}
\email{xevra86@gmail.com}

\author[0009-0008-5622-6857]{Emily McPike}
\affiliation{Department of Astrophysics, American Museum of Natural History, New York, NY 10024, USA}
\affiliation{Graduate Center, City University of New York, 365 5th Avenue, New York, NY 10016, USA}
\email{emcpike@gradcenter.cuny.edu}

\author[0000-0003-0738-8186]{Jake Postiglione}
\affiliation{Department of Astrophysics, American Museum of Natural History, New York, NY 10024, USA}
\affiliation{Graduate Center, City University of New York, 365 5th Avenue, New York, NY 10016, USA}
\email{jpostiglione@gradcenter.cuny.edu}

\author[0000-0001-7163-8712]{Harrison E. Cook}
\affiliation{New Mexico State University, Department of Astronomy, PO Box 30001 MSC 4500, Las Cruces, NM 88001, USA}
\email{hecook@nmsu.edu}

\author[0000-0003-2430-9515]{Kaila Nathaniel}
\affiliation{Center for Computational Relativity and Gravitation, Rochester Institute of Technology, Rochester, New York 14623, USA}
\email{kjn9240@rit.edu}

\author[0009-0004-4600-5074]{Nicolas Posner}
\affiliation{American Museum of Natural History, 200 Central Park West, New York, NY 10024, US}
\email{nicolasposner@gmail.com}

\author[0000-0002-0786-7307]{Barry McKernan}
\affiliation{Center for Computational Astrophysics, Flatiron Institute, 
162 5th Ave, New York, NY 10010, USA}
\affiliation{Department of Astrophysics, American Museum of Natural History, New York, NY 10024, USA}
\affiliation{Department of Science, BMCC, City University of New York, New York, NY 10007, USA}
\affiliation{Graduate Center, City University of New York, 365 5th Avenue, New York, NY 10016, USA}
\email{bmckernan@amnh.org}

\author[0000-0002-5956-851X]{K.E. Saavik Ford}
\affiliation{Center for Computational Astrophysics, Flatiron Institute, 
162 5th Ave, New York, NY 10010, USA}
\affiliation{Department of Astrophysics, American Museum of Natural History, New York, NY 10024, USA}
\affiliation{Department of Science, BMCC, City University of New York, New York, NY 10007, USA}
\affiliation{Graduate Center, City University of New York, 365 5th Avenue, New York, NY 10016, USA}
\email{saavikford@gmail.com}

\author[0000-0001-5832-8517]{Richard O'Shaughnessy}
\affiliation{Center for Computational Relativity and Gravitation, Rochester Institute of Technology, Rochester, New York 14623, USA}
\email{rossma@rit.edu}

\begin{abstract}
\label{abstract}
    \noindent Binary black hole mergers (BBH)  are currently observed by the gravitational-wave detectors LIGO, Virgo, and KAGRA (LVK). A fraction of these mergers are likely to originate in active galactic nuclei (AGN), but the details of the dynamical processes driving AGN channel  mergers remain poorly understood. Monte Carlo For AGN Channel Testing and Simulation (McFACTS) is an open-source, public software used to simulate AGN channel BBH formation and evolution with configurable options for exploring various AGN and Nuclear Star Cluster (NSC) parameter options. Within \texttt{McFACTS}, BBH mergers evolve through gas drag and dynamical interactions, ignoring relativistic effects which can influence merger remnant properties. Here we demonstrate in \texttt{McFACTS} the effect of handing off inspiralling BBH to a numerical relativity (NR) surrogate model, which more accurately predicts the masses, spins, and kick velocities of merger remnants. Compared to a simple analytic approach to calculating merger and remnant properties the NR surrogate model is significantly more accurate at predicting remnant spins, which improve $\chi_{\rm eff}$ and $\chi_{\rm p}$ calculations, and recoil kicks, parameters especially  important for higher generation BBH mergers. Our findings show the NR surrogate significantly changes distributions of remnant spins and kick velocities, as well as showing that alternative methods should be used cautiously in future studies. We also outline some simple models which are computationally efficient and can reproduce much of the NR surrogate model results. 

\end{abstract}


\section{introduction}
The binary black hole (BBH) mergers observed in gravitational waves \cite{gwtc5_catalog, gwtc5_pop} are thought to originate from multiple distinct formation channels \citep{zevin2021, cheng2023, storm2025}. Among them are the isolated formation of stellar binaries \citep[e.g.][]{delfavero2023, eldridge2017, breivik2020, fragos2023, levina2026}, dynamic formation in dense cluster environments \citep[e.g.][]{ye2026, fishbach_holz_farr2017, rodriguez2016, ye_fishbach2024, gerosa_fishbach2021, rinaldi2026, li2026}, and dynamic formation in the accretion disks of AGN \citep[e.g.][]{McK2014,Bartos17,Stone17,Tagawa2020,gayathri2022}. Repeated or hierarchical BBH mergers that form BHs of higher masses and spins can occur either in dense stellar environments, or in deep potential wells, or both. Hierarchical merger model efficiency therefore depends on three things: the magnitude of the kick that the BBH merger remnant experiences, the density of BH in the local environment, and most importantly, the local escape velocity. Therefore, the details of the kick matter in determining where a BBH merger could have occurred.

When two BHs merge and emit gravitational waves (GWs), part of their linear momentum is imparted back onto the merger remnant due to the linear momentum carried by the radiation, which generates a gravitational recoil, or kick, onto the final merger remnant \citep{fichett1983, gonzalez2007}. The magnitude and direction of the kick depend on the merging BBH's mass ratio, spin magnitudes and spin alignments. High, or even modest, kick magnitudes lead to a very low retention rate of merged BHs (and low hierarchical merger efficiency) unless the local escape velocity is very high. This is why BBH mergers in active galactic nuclei (AGN) are expected to be more hierarchically efficient than any other astrophysical environment \citep[e.g. for a review see][]{ford_using_2025}, and so a very promising place for the highest mass  (and spin) BBH mergers currently observed by LVK \citep[e.g.][]{gw231123,delfavero_prospects_2025}. 
Nevertheless, even in the AGN channel, the details of kicks matter. Lower magnitude kicks with in-plane directions allow merged BHs to be retained within the disk during an AGN episode. This influences the efficiency of hierarchical mergers during a given AGN episode and also influences potential electromagnetic counterparts \citep{mckernan_ram-pressure_2019,graham_candidate_2020,tagawa_observable_2023,McPike2026}.

Here we discuss the effect on AGN channel distributions of replacing approximate analytic BBH merger prescriptions with far more accurate numerical relativity (NR) surrogate models of the final, highly relativistic stages of BBH merger.


\section{methods}
\label{sec:intro_rotation}
We use the the public, open-source code: Monte Carlo for AGN channel testing and simulation (\texttt{McFACTS}) \citep{mckernan_mcfacts_2025} to model the AGN channel. This section describes how BBH merger remnant properties are calculated in \texttt{McFACTS}. 

\subsection{\texttt{McFACTS} Default}

\subsubsection{\texttt{McFACTS} Structure}
In this work we used \texttt{McFACTS v0.4.1}. 
\texttt{McFACTS} takes three basic inputs: the supermassive black hole (SMBH) mass, the nuclear star cluster (NSC) parameters and disk model parameters. The initial conditions for these inputs are set in a file modified by the user (see Table \ref{tab:default} for \texttt{model$\_$choice.ini}), along with a random number seed (\texttt{SEED}). When simulating multiple galaxies with the same input parameters, \texttt{SEED} iterates by one for each galaxy, ensuring reproducibility for any given \texttt{SEED} and \texttt{.ini} file.

In this work we compare four models (see below for details) of the final stage of BBH mergers: 1) \texttt{McFACTSnosur}, the default analytic treatment based on \citep{TichyM2008}, 2) \texttt{McFACTSsur} a NR surrogate model based on \textit{surfinBH} \citep{varma2019}, 3) \texttt{McFACTSprec}, a model based on \texttt{precession}  \citep{precession} and 4) \texttt{McFACTSfilter}, a simple numerical filter based on  merger details. We choose between these models in \texttt{model$\_$choice.ini} using two new boolean parameters:  \texttt{flag$\_$use$\_$surrogate} and \texttt{flag$\_$use$\_$spin$\_$check}. 
Setting \texttt{flag$\_$use$\_$surrogate}$=[-1,0,1]$ selects methods [\texttt{McFACTSnosur} \texttt{McFACTSsur}, \texttt{McFACTSprec}] and \texttt{flag$\_$use$\_$surrogate}$=0$, \texttt{flag$\_$use$\_$spin$\_$check}$=1$ selects \texttt{McFACTSfilter}. Here we modified default values from \citep{McFACTS_I} to (new values) of \texttt{timestep$\_$num = 100(60)},  \texttt{galaxy$\_$num = 100(1000)}, and  \texttt{fraction$\_$bin$\_$retro = 0.0(0.5)} (see Table~\ref{tab:default}).

\begin{table}[]
    \footnotesize
    \centering
    \caption{Slightly modified from \cite{mckernan_mcfacts_2025}, these are the default input parameters for the SMBH, AGN disk, and NSC as well as physics choices in \texttt{model$\_$choice.ini}. We do not include initialization of star parameters since stars are turned off by default in \texttt{v.0.3.0}.}
    \label{tab:default}
    \begin{tabular}{c|c|c}
\hline
Parameter & Default Value & Units \\
\hline
\texttt{disk$\_$model$\_$name} & \texttt{sirko$\_$goodman} & \\
\texttt{flag$\_$use$\_$pagn} & 1 & \\
\texttt{flag$\_$use$\_$surrogate} & 0 & \\
\texttt{flag$\_$use$\_$spin$\_$check} & 0 & \\
\texttt{flag$\_$phenom$\_$turb} & 0 & \\
\texttt{phenom$\_$turb$\_$centroid} & 0 & \\
\texttt{phenom$\_$turb$\_$std$\_$dev} & 1.0 & \\
\texttt{disk$\_$radius$\_$trap} & 700 & $\rm{r}_{g,\rm{SMBH}}$ \\
\texttt{disk$\_$radius$\_$outer} & $5\times 10^{4}$ & $\rm{r}_{g,\rm{SMBH}}$ \\
\texttt{disk$\_$radius$\_$max$\_$pc} & 0 & \\
\texttt{disk$\_$alpha$\_$viscosity} & 0.01 &  
\\
\texttt{inner$\_$disk$\_$outer$\_$radius} & $50$ & $\rm{r}_{g,\rm{SMBH}}$ \\
\texttt{disk$\_$inner$\_$stable$\_$circ$\_$orb} & $6$ & $\rm{r}_{g,\rm{SMBH}}$ \\
\texttt{disk$\_$aspect$\_$ratio$\_$avg} & $0.03$ &  \\
\hline
\texttt{nsc$\_$radius$\_$outer} & 5 & pc\\
\texttt{nsc$\_$radius$\_$crit} & 0.25 & pc\\
\texttt{nsc$\_$mass} & $3 \times 10^{7}$ & $M_{\odot}$ \\
\texttt{nsc$\_$ratio$\_$bh$\_$num$\_$star$\_$num} & $10^{-3}$ &  \\
\texttt{nsc$\_$ratio$\_$bh$\_$mass$\_$star$\_$mass} & 10 &  \\
\texttt{nsc$\_$density$\_$index$\_$inner} & 1.75 & \\
\texttt{nsc$\_$density$\_$index$\_$outer} & 2.5 & \\
\texttt{nsc$\_$spheroid$\_$normalization} & 1 &  \\
\texttt{nsc$\_$imf$\_$bh$\_$mode} & 10 & $M_{\odot}$ \\
\texttt{nsc$\_$imf$\_$bh$\_$max} & 40 & $M_{\odot}$ \\
\texttt{nsc$\_$imf$\_$powerlaw$\_$index} & 2 & \\
\texttt{nsc$\_$imf$\_$bh$\_$method} & \texttt{default} & \\
\texttt{nsc$\_$bh$\_$spin$\_$dist$\_$mu} & 0 & \\
\texttt{nsc$\_$bh$\_$spin$\_$dist$\_$sigma} & 0.1 & \\
\texttt{nsc$\_$imf$\_$powerlaw$\_$index} & 2 & \\
\hline
\texttt{disk$\_$bh$\_$torque$\_$condition} & 0.1 & \\
\texttt{disk$\_$bh$\_$eddington$\_$ratio} & 1.0 & \\
\texttt{disk$\_$bh$\_$orb$\_$ecc$\_$max$\_$init} & 0.9 & \\
\texttt{disk$\_$radius$\_$capture$\_$outer} & $10^{3}$ & $\rm{r}_{g,\rm{SMBH}}$ \\
\texttt{disk$\_$bh$\_$pro$\_$orb$\_$ecc$\_$crit} & 0.01 & \\
\texttt{mass$\_$pile$\_$up} & 35 & $M_{\odot}$\\
\texttt{timestep$\_$duration$\_$yr} & $10^{4}$ & yr\\
\texttt{timestep$\_$num} & 60 & \\
\texttt{capture$\_$time$\_$yr} & $10^{5}$ & yr\\
\texttt{galaxy$\_$num} & 1,000& \\
\texttt{delta$\_$energy$\_$strong$\_$mu} & 0.1 & \\
\texttt{delta$\_$energy$\_$strong$\_$sigma} & 0.02 & \\
\texttt{fraction$\_$bin$\_$retro} & 0.0 & \\
\texttt{flag$\_$thermal$\_$feedback} & 1 & \\
\texttt{flag$\_$orb$\_$ecc$\_$damping} & 1 & \\
\texttt{flag$\_$dynamic$\_$enc} & 1 & \\
\texttt{flag$\_$torque$\_$prescription} & \texttt{paardekooper} & \\
\texttt{harden$\_$energy$\_$delta$\_$mu}  & 0.9 & \\
\texttt{harden$\_$energy$\_$delta$\_$sigma} & 0.025 & \\
\hline
    \end{tabular}
\end{table}

Note there are many analytical approximations of BBH merger remnant mass in the literature \citep[e.g.][]{TichyM2008,lousto2010,pann2013,lousto_zloch2014,lz2016}. Of these,  \cite{TichyM2008} (T\&M08) was the approach initially used in \texttt{McFACTS} for its ease of applicability as well as its computational efficiency. 

\subsubsection{Mass}

The mass of a remnant BH in \texttt{McFACTSnosur} is given by \citep{TichyM2008}:
\begin{equation}
    m=1+(m^0 - 1)4\eta + m^{a1}16\eta^2(a_z+b_z)
\end{equation}
where $m$ is the final mass of the binary. The linear order coefficients are $m^0=0.9515$ and $m^{a1}=-0.13$. The coefficient $\eta$ is the symmetric mass ratio: $\eta = \frac{q}{(1+q)^2}$, where $q\equiv m_{2}/M_{1}$ is the mass ratio with $m_{2}$ and $M_{1}$ the masses of the secondary and primary BHs. The component of the progenitor spins along the direction of the orbital angular momentum are represented as $a_z$ and $b_z$ and are initially drawn from a Gaussian distribution centered on $\chi =0$ with a variance $\sigma_{\chi}=0.2$ (\cite{mckernan_mcfacts_2025}).

\subsubsection{Spin}

Black holes have intrinsic spins presumably determined by how they formed from progenitor stars, and these spins can change orientation during the BBH coalescence \citep[e.g.][]{lousto2010}. The dominant component of the final spin of the remnant is set by the total angular momentum of the BBH at merger. To conserve angular momentum in the system, the excess is carried away by GWs  (\cite{bek1973}). The spin of the remnant depends on the BBH mass ratio, $q=m_{2}/M_{1}$ and the initial spin magnitudes and orientations of the inspiralling BHs. Accurately calculating the spin (by comparing the four models listed above) allows for better comparisons to LVK observations, and ensures subsequent hierarchical mergers have accurate initial spins for higher generation progenitors.

In \texttt{McFACTS} the vector component spins of each BH are tracked and are used to calculate $\chi_{\rm{eff}}$ and $\chi_{\rm{p}}$, the two best measured spin parameters measured by \cite{gwtc5_pop, gwtc4pop}. 
The mass weighted spin of the BBH merger is given by \cite{damour2001}:
\begin{equation}
    \chi_{\rm{eff}} = \frac{M_1 \overrightarrow{\chi_1} + m_2 \overrightarrow{\chi_2}}{M_1 + m_2} \ \cdot \ \hat{L}_{\rm{bin}}
\end{equation}
where $\hat{L}_{\textrm{bin}}$ is the BBH orbital angular momentum around its own center of mass. 

The effective precession spin component of the binary is calculated as \citep{schmidt2015}:
\begin{equation}
    \chi_p = \frac{1}{A_2m_2^2} \textnormal{max} \left[ A_1S_{1\perp}, A_2, S_{2\perp} \right]
\end{equation}

\noindent where the constants $A_1=2+3q/2$ and $A_2=2+3/(2q)$ and $q=m_{2}/M_{1}$. 

The remnant spin equations applied to the BH populations in default \texttt{McFACTS} (model \texttt{McFACTSnosur} above) come from T\&M08 and are described as three component equations in the coordinate system such that z-axis is perpendicular to the initial orbital plane. The binary center of mass is initially at rest at the origin, with the x-axis along direction of the momentum of BH b, and the y-axis along the line connecting the BHs with BH a located at y $>$ 0. The component spins are as follows:
\begin{equation}
    s_x = s_x^{a1} [g(q, c)a_x + g(1/q, c)b_x] - s_y^{a1}4\eta(a_y+b_y)
\end{equation}
\begin{equation}
    s_y = s_y^{a1} 4\eta(a_x+b_x) + s_x^{a1}[g(q, c)a_x + g(1/q, c)b_x]
\end{equation}

\begin{multline} \label{eq:s_z}
    s_z = s_z^{(0)} (4\omega\eta + 16(1-\omega)\eta^2) \\ + 
    s_z^{a1}[g(q, c_3)a_z + g(1/q, c_3)b_z] \\ + 
    16k\eta^2[(a_x+b_x)^2+(a_y+b_y)^2-(a_z+b_y)^2] 
\end{multline}
\\
where $a_x$, $b_x$ are the x-components of the progenitor spins and, $a_y$, $b_y$, are the y-components of the progenitor spins. The linear order coefficient values are listed in Table~\ref{tab:spin_coeff} and the functions are represented by: 
\begin{equation}
    g(q,c) = \frac{(c+1)^2}{(c+q)^2}, \qquad
    g(1/q,c) = \frac{(c+1)^2}{(c+\frac{1}{q})^2}
\end{equation}
\begin{equation}
    c = \frac{\sqrt{s_x^{a1}}}{1-\sqrt{s_x^{a1}}} = 0.762, \qquad
    c_3 = \frac{\sqrt{s_z^{a1}}}{1-\sqrt{s_z^{a1}}} = 0.632
\end{equation}

\begin{table}[]
    \centering
    \begin{tabular}{cc}\toprule
        \textbf{Coefficient}&\textbf{Value}\\
        \midrule
        $s_x^{a1} = s_x^{b1}$&0.187\\
        $s_y^{a1} = s_y^{b1}$&0.028\\
        $s_z^{a1} = s_z^{b1}$&0.15\\
        $s_z^0$&0.686\\
 $\omega$&1.26\\
 $k$&0.008\\ \bottomrule\end{tabular}
    \caption{Linear coefficients for the x-y-z components of the merger remnant spin, T\&M08}
    \label{tab:spin_coeff}
\end{table}

The spin components, $s_x, s_y, s_z$, are used to determine a dimensionless spin, $s$, but in this work we denote the dimensionless spin as $a$.

The approximations underlying T\&M08 guarantee accurate remnant mass and spin over a narrow range of mass ratios, $q=[0.67, 1.0]$. But random mergers drawn from the \texttt{McFACTS} default IMF span mass ratios $q=[0.20,1.0]$, and higher generation mergers can be even more extreme ($q \leq 0.1$) . Since hierarchical mergers may frequently involve mergers with mass ratios outside of the validated range appropriate to T\&M08, and hierarchical mergers are a key component of AGN channel populations, we expect that our default approximation (\texttt{McFACTSnosur}) should break down compared to a more accurate NR surrogate model (\texttt{McFACTSsur}), particularly for higher generation, small $q$, mergers. 

\subsubsection{Kick Velocity}

The default calculation (\texttt{McFACTSnosur}) for calculating the kick velocity associated with a BBH merger is given by  \citep{lousto2010, lousto2012}
\begin{equation}
    v_{kick} = v_m\, \hat{x} + v_\perp (\cos{\xi}\,\hat{x} + \sin{\xi}\,\hat{y}) + v_\parallel\, \hat{z}
    \label{eq:kick}
\end{equation}
where $v_m$, $v_\parallel$, and $v_\perp$ are represented by
\begin{equation}
    v_m = A\eta^2\sqrt{1-4\eta} (1+B\eta) 
\end{equation}
\begin{equation}
        v_\parallel = \frac{H\eta^2}{1+q} (\chi_{2, \parallel} - q\chi_{1, \parallel})
\end{equation}
\begin{multline}
    v_{\perp} = \frac{16\eta^2}{1+q} [V_{1,1} + V_A \tilde{S}_\parallel + V_B \tilde{S}_\parallel^2 +V_C \tilde{S}_\parallel^3] \\ 
    \times |\chi_{2,\perp} - q\chi_{1, \perp} | \cos({\phi_\Delta} - \phi_1)
    \label{eq:v_perp}
\end{multline}
and the subscripts $_1, _2$ represent the primary and secondary components and $\chi_1$ and $\chi_2$ represent the progenitors' dimensionless spin vectors. The coefficient values are listed in Table~\ref{tab:kick_const}. The vector $\tilde{S}$ is represented by
\begin{equation}
    \tilde{S} = 2 \frac{\chi_2 + q^2\chi_1}{(1+q)^2}
\end{equation}
where $\phi_1$ is represented as the phase angle of the binary and $\phi_{\Delta}$ is the angle between the in-plane spin angle component of
\begin{equation}
    \Delta = M^2 \frac{\chi_2 - q\chi_1}{1+q} 
    \label{eq:delta}
\end{equation}
and the infall direction at merger. Since $\phi_{\Delta} -\phi_1$ is treated as a random distribution in $[0,2\pi)$, Eq. \ref{eq:delta} is not used in calculations. Note, since recoil kick velocity depends on the spin vector of each progenitor, deviations in the spin calculations for each merger can have far reaching effects on hierarchical mergers.

This default method (\texttt{McFACTSnosur}) predicts recoil velocities within $\sim 8\%$ of non-linear numerical simulations down to $q \sim 0.1$. 

\begin{table}[]
    \centering
    \begin{tabular}{cc}\toprule
         \textbf{Coefficients}&  \textbf{Value}\\\midrule
         A& 
    $1.2\times10^4 \ \text{km s}^{-1}$\\
 B&$-0.93$\\
 H&$6.9 \times 10^3 \ \text{km s}^{-1}$ \\
 $\xi$&$145^{\circ}$\\
 $V_{1,1}$&$3678 \ \text{km s}^{-1}$ \\
 $V_A$&$2841 \ \text{km s}^{-1}$\\
 $V_B$&$1793  \ \text{km s}^{-1}$ \\
 $V_C$&$1507  \ \text{km s}^{-1}$ \\ \bottomrule\end{tabular}
    \caption{Coefficients applied to Eq. \ref{eq:kick}$-$\ref{eq:v_perp} as well as Eq. \ref{eq:v_k_gerosa}$-$Eq.\ref{eq:v_s_gerosa}}
    \label{tab:kick_const}
\end{table}

\subsection{Numerical Relativity Surrogate}
\label{NRsur}

The most accurate approach for calculating BBH merger dynamics for arbitrary progenitor parameters according to general relativity is numerical relativity. However, NR methods are computationally expensive and require immense resources when applied to large populations of mergers such as those in  \texttt{McFACTS}. Surrogate models have been created to reduce the computational resources needed to obtain merger and remnant parameters for arbitrary systems, without direct numerical evolution.

Surrogate models describing BH mergers save computational expense by interpolating the outputs of fully evolved numerical solutions over the input parameter space to build a map of the output remnant parameters. The NR surrogate model is a state-of-the-art numerical model trained on numerical relativity simulations using Gaussian process regression to predict the merger remnant's mass, spin, and kick velocity as a function of the progenitor parameters. The NR surrogate model used in \texttt{McFACTS} (our model \texttt{McFACTSsur}) is a modified version of \textit{surfinBH}, made by \cite{varma2019}, which has been trained on simulations from the SXS Catalog down to mass ratio $q\geq0.125$~\cite{Scheel:2025jct}.\footnote{This remnant surrogate will be made public and an analysis of its accuracy will be presented in an upcoming publication.} 

To integrate the NR surrogate into \texttt{McFACTS}, we take BBHs flagged as merged and pass the binary parameters from the timestep just before the merger flag to the remnant prediction pipeline. These parameters are the masses, spin magnitudes, spin and precession angles with respect to the binary orbital angular momentum, the inclination of the binary, the mass of the SMBH, and the spin of the SMBH. With these parameters \texttt{McFACTSsur} then evolves the binary according to post-Newtonian (PN) theory at 3.5 PN order with the TaylorT1 approximant~\cite{Boyle_PostNewtonian_jl_2024}, until the PN equations break down, which we use as a proxy for the time of peak GW strain. The binary parameters are then extracted from the PN evolution 100$M$ (for $M$ the total mass of the binary) before this break down point and fed into the NR surrogate, which requires input parameters defined $100M$ before the time of peak GW strain. With this information, the NR surrogate provides a prediction of the remnant quantities. This remnant information is then passed back to \texttt{McFACTS} for further dynamical evolution within the disk and nuclear star cluster system.

\subsection{Alternative Methods}
The following section describes the alternative methods used to calculate the spin and kick velocity of the final merger remnant (models \texttt{McFACTSprec} and \texttt{McFACTSfilter}).

\subsubsection{\texttt{precession}}
\label{McFACTSprec}

An alternative analytical approach to BBH remnant properties is the \texttt{precession} package \citep{precession}. \texttt{precession} applies PN fitting formulas to calculate BBH remnant properties and is designed  to be used in the context of large-scale numerical studies such as population synthesis for GW  sources and we implement it here as \texttt{McFACTSprec}. While \texttt{McFACTSsur} evolves the BBH inspiral with PN methods up to $100M$ before the time of peak GW strain before passing the calculations off to the NR surrogate, \texttt{McFACTsprec} calculates the BBH inspiral with PN methods up to $10M$ before using a standard integrator of the orbit-averaged PN equations to complete the BBH inspiral. 

The final remnant mass is \citep{precession}
\begin{multline}
    \frac{M_f}{M} = 1 - \eta (1-4\eta)(1-E_{ISCO}) \\ - 16\eta^2[p_0+4p_1\tilde{\chi_{\parallel}}(\tilde{\chi}_{\parallel}+1)]
    \label{eq:mass_gerosa}
\end{multline}
where $M_f$ is the mass of the BH remnant, $M$ is the binary's total mass, $\tilde{\chi}_{\parallel}$ is the parallel spin projection, and $\eta = \frac{q}{(1+q)^2}$, where $q\equiv m_{2}/M_{1}$ is the mass ratio of the secondary and primary BH. The parameters $p_0 = 0.04827$ and $p_1 = 0.01707$ are generated by fitting 186 numerical relativity simulations from various groups (\cite{barausse2012}). The parallel spin projection, and other similar values, are found using a weighted combination of the BH spin:
\begin{equation}
\boldsymbol{\Delta} = \frac{\chi_1 \boldsymbol{\hat{S_1}} - q\chi_2 \boldsymbol{\hat{S_2}}}{1+q}, \ \ \ \ 
\boldsymbol{\tilde{\chi}} = \frac{q^2\chi_2 \boldsymbol{\hat{S_2}} + \chi_1 \boldsymbol{\hat{S_1}}}{(1+q)^2}
\end{equation}
and their projections parallel or perpendicular to the orbital angular momentum: $\tilde{\chi}_{\parallel} = \boldsymbol{\tilde{\chi}} \cdot \boldsymbol{\hat{L}}$, $\tilde{\chi}_{\perp} = | \boldsymbol{\tilde{\chi}} \times \boldsymbol{\hat{L}}|$, $\Delta_{\parallel} = \boldsymbol{\Delta} \cdot \boldsymbol{\hat{L}}$, $\Delta_{\perp} = | \boldsymbol{\Delta} \times \boldsymbol{\hat{L}} |$. 

The vectors $\hat{\textbf{S}}$ and $\hat{\textbf{L}}$ are the directions of the spin of the BHs and the angular momentum w.r.t. the binary's orbital plane. The $E_{ISCO}$ is the energy per unit mass of an effective particle of spin, $ \boldsymbol{\tilde{\chi}}$, at the innermost stable circular orbit:
\begin{equation}
    E_{ISCO} = \sqrt{1-\frac{2}{3r_{ISCO}}}
\end{equation}

\begin{multline}
    r_{ISCO} = 3 + Z_2 \\ 
    - \mathrm{sign}(\tilde{\chi}_{\parallel})\sqrt{(3-Z_1)(3+Z_1+2Z_2)}
\end{multline}

\begin{equation}
    Z_1 = 1+(1-\tilde{\chi}_{\parallel}^2)^{1/3} [(1+\tilde{\chi}_{\parallel})^{1/3}+(1-\tilde{\chi}_{\parallel})^{1/3}]
\end{equation}

\begin{equation}
    Z_2 = \sqrt{3\tilde{\chi}_{\parallel}^2 + Z_1^2}
\end{equation}

\noindent The final remnant spin is calculated by:
\begin{equation}
    \chi_f = \mathrm{min} \left(1, \left| \boldsymbol{\tilde{\chi}} + \frac{q}{(1+q)^2} \ell \boldsymbol{\hat{L}}\right| \right)
\end{equation}

\begin{multline}
    \ell = 2\sqrt3 + t_2\eta + t_3\eta^2 \\ 
    + s_4 \frac{(1+q)^4}{(1+q^2)^2} \tilde{\chi}^2 + (s_5\eta + t_0 +2) \frac{(1+q)^2}{1+q^2}\tilde{\chi}_{\parallel}
\end{multline}

\noindent The final remnant kick velocity magnitude is calculated by:
\begin{equation}
    v_k = \sqrt{v_m^2 + 2v_m v_{s\perp} \cos{\zeta} + v_{s\perp}^2 + v_{s\parallel}^2}
    \label{eq:v_k_gerosa}
\end{equation}

\noindent where $\zeta$ is the angle between the velocity components of the mass term, $v_m$, and the orbital-plane spin term, $v_{s\perp}$, defined by:
\begin{equation}
    v_m = A\eta^2 \frac{1-q}{1+q} (1+B\eta)
\end{equation}

\begin{equation}
    v_{s\perp} = H\eta^2 \Delta_{\parallel}
\end{equation}

\begin{multline}
    v_{s\parallel} = 16\eta^2 [\Delta_{\perp}(V_{11}+2V_A\tilde{\chi}_{\parallel} + 4V_B\tilde{\chi}_{\parallel}^2 + 8V_C\tilde{\chi}_{\parallel}^3) \\ 
    + 2\tilde{\chi}_{\perp}\Delta_{\parallel}(C_2+2C_3\tilde{\chi}_{\parallel})] \cos{\Theta}
    \label{eq:v_s_gerosa}
\end{multline}

\noindent where $\Theta$ is the initial separation of the binary and is assumed to be a random number from a uniform distribution of $[0,\pi)$. The coefficients match those used in \cite{TichyM2008} and are listed in Table \ref{tab:kick_const}. The coefficients used only by \texttt{precession} \cite{gerosa_kesden2016} are listed in Table \ref{tab:gerosa_coeff}. 

\begin{table}[]
    \centering
    \begin{tabular}{ccc}\toprule
          \textbf{Parameter}&\textbf{Coefficients}&  \textbf{Value}\\\midrule
          \underline{Mass}&$p_0$& 
    $0.04827$\\
  &$p_1$&$0.01707$\\
  \underline{Spin}&$t_0$&$-2.8904$\\
  &$t_2$&$-3.51712$\\
  &$t_3$&$2.5763$\\
  &$s_4$&$-0.1229$\\
  &$s_5$&$0.4537$\\
  \underline{Kick Velocity}&$C_2$&$1140 \ \text{km s}^{-1}$\\ 
 & $C_3$&$2481 \ \text{km s}^{-1}$\\ \bottomrule\end{tabular}
    \caption{Coefficients applied to Eq. \ref{eq:mass_gerosa}$-$Eq.\ref{eq:v_s_gerosa}}
    \label{tab:gerosa_coeff}
\end{table}

\subsubsection{Hard Coded Filter}
\label{McFACTSfilter}

Strictly to minimize computational expense, we have also implemented a simple numerical filter method ( \texttt{McFACTSfilter}) for generating distributions of remnant properties that are more accurate than \texttt{McFACTSnosur}. \texttt{McFACTSfilter} takes higher generation mergers from the \texttt{McFACTSnosur} model and resets their calculated $a_{final}$ to a value selected from a random distribution within bounds set by our results from surrogate simulations. This hard filter method is not recommended for calculations that require exact values of $a_{final}$; it has been made purely on the basis of minimizing computational expense.

As we will see below, we found good agreement between T\&M08 and NR results for first generation ($1g-1g$) mergers, so we found no need to filter first generation mergers, where $1g(2g)=$first (second) generation and indicates BH that have zero(one) prior mergers.  For $2g-mg$ $(m\leq2)$ mergers, where \texttt{McFACTSnosur} finds $a_{final}<0.7$,  \texttt{McFACTSfilter} chooses $a_{final}$ randomly from a uniform distribution $a_{f}=[0.75, 0.85]$. For $3g-ng$ $(n\leq 3)$ mergers, where \texttt{McFACTSnosur} finds $a_{final}<0.85$, \texttt{McFACTSfilter} chooses $a_{final}$ randomly from a uniform distribution $a_{final}=[0.85, 0.95]$. Thus, $a_{final}$ values lie within the range of final spins found by the NR surrogate for similar mergers.


\section{Results}

This section describes and compares  the results of each method used in calculating BBH remnant parameters for the input parameters listed in Table \ref{tab:default}.  In each of the plots below, gold circles, purple downward pointing triangles and red upward pointing triangles represent $1g,2g,\geq3g$ merger remnants respectively.

\subsection{Spin}

\begin{sidewaysfigure*}
    \centering
    \includegraphics[width=\linewidth]{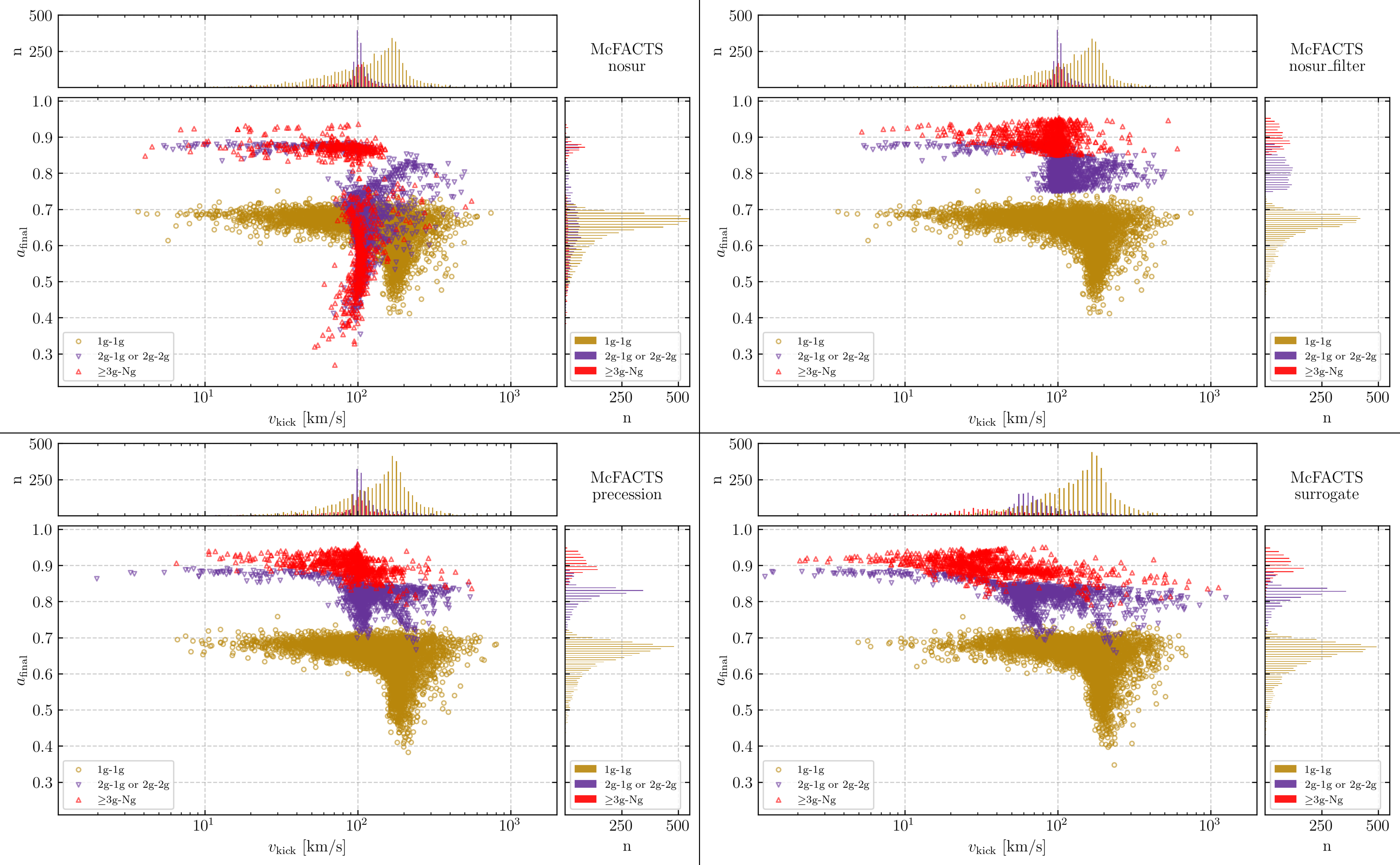}
    \caption{Spin vs kick with histogram distributions of each parameter}
    \label{fig:spin_v_kick}
\end{sidewaysfigure*} 

Figure~\ref{fig:spin_v_kick} compares $a_{\mathrm{final}}$ as calculated by our four models (clockwise from top-left: \texttt{McFACTSnosur}, \texttt{McFACTSfilter}, \texttt{McFACTSsur}, \texttt{McFACTSprec}).The central panel in each case shows the distribution of $a_{\mathrm{final}}$ as a function of the remnant kick velocity, $v_{\mathrm{kick}}$, for each generation. The right (top) panel in each case shows the number distribution of $a_{\mathrm{final}}$($v_{\mathrm{kick}}$) respectively. 

The results of the most physically accurate model in Fig.~\ref{fig:spin_v_kick} are in the bottom right panel (
\texttt{McFACTSsur}). Among the 1g mergers (gold points), these results look very similar to the other three models. However, the distribution of $a_{\mathrm{final}}$ for higher generation (purple, red) mergers is quite different particularly when compared to our default \texttt{McFACTSnosur} (top left), as we expected. In particular: higher generation mergers (purple, red) have a wider range of  $v_{\mathrm{kick}} \sim [1,2000]{\rm km/s}$ with \texttt{McFACTSsur} than with any other model, there is a  bimodal distribution in $v_{\rm kick}$ among the $2g$ (purple) mergers, and there is a relatively clear separation in $a_{\rm final}$ between the merger generations. Evidently we should not rely on \texttt{McFACTSnosur} (the default model) for determining remnant spins and kick velocities of higher generation (purple, red) mergers. The bottom left panel, showing results using \texttt{McFACTSprec} is similar in form to those for \texttt{McFACTSsur} but with narrower distributions in $v_{\rm kick} \sim [20,600]{\rm km/s}$ for higher generations.

In order to quantify the reliability of 
compare each model to \texttt{McFACTSsur}, we used a Kolmogorov–Smirnov (KS) Test to generate KS Statistics and p-values for the $v_{\mathrm{kick}}$ and $a_{\mathrm{final}}$ for each generation of mergers for each model.

\subsubsection{\texttt{McFACTSnosur}} 

For \texttt{McFACTSnosur} in Fig.~\ref{fig:spin_v_kick}, the $1g$ mergers (gold points) form a distribution of spin values peaking at $a_{\rm final} \sim 0.67$ and plateauing around $a_{\rm final} \sim 0.70$. $1g$ mergers with  $a_{\mathrm{final}}=[0.6, 0.7]$, span  a wide range of $v_{\mathrm{kick}}=[5,600]$ km/s, while $1g$ mergers with $a_{\mathrm{final}} \leq 0.6$ have a narrow distribution of  $v_{\mathrm{kick}} \sim 200 \mathrm{km/s}$. The $2g$ mergers (purple) form a broad distribution across $a_{\rm final} \sim [0.35, 0.9]$. For $2g$ mergers with $a_{\mathrm{final}} < 0.85$, $v_{\mathrm{kick}} \sim 100{
\rm km/s}$. For $2g$ mergers with $a_{\mathrm{final}} > 0.85$, $v_{\mathrm{kick}}\sim [5, 100]\mathrm{km/s}$. These trends in the spin distribution are also found for $\geq 3g$ mergers (red). The right sub-panel shows the distribution by number in  $a_{\mathrm{final}}$, illustrating that higher generation mergers (purple and red) are relatively uniform across $a_{\mathrm{final}}$ except at the highest spin values. 

As shown in Table \ref{tab:kstest}, the p-values of $a_{\mathrm{final}}$ for each merger generation are less than the commonly used significance level, 0.05. This tells us none of the $a_{\mathrm{final}}$ values calculated for \texttt{McFACTSnosur} match those calculated for \texttt{McFACTSsur}, indicating they should not be used when investigating hierarchical merger spin populations.

\begin{table}
    \centering
    \begin{tabular}{cccc}\toprule
         \multicolumn{4}{c}{\textbf{KS Test}}\\\midrule
         \textbf{Method}&  \textbf{Generation}& \textbf{KS Statistic}& \textbf{P-Value}\\\midrule
         \multicolumn{4}{c}{$\boldsymbol{a_{final}}$}\\\midrule
         \texttt{\underline{McFACTSnosur}}&  $1g$&  0.03& $\approx 0.0$\\
         &  $2g-ng$&  0.76& $\approx 0.0$\\
         &  $\geq 3g$&  0.71& $\approx 0.0$\\
         \texttt{\underline{McFACTSfilter}}&  $1g$&  0.04& $\approx 0.0$\\
         &$2g-ng$& 0.27& $\approx 0.0$\\
         &  $\geq 3g$&  0.12& $\approx 0.0$\\
         \texttt{\underline{McFACTSprec}}&  $1g$&  0.01& 0.94\\
         & $2g-ng$& 0.01& 0.99\\
         & $\geq 3g$& 0.04& 0.30\\
         \texttt{\underline{McFACTSsur}}& $1g$& 0.0& 1.0\\
         & $2g-ng$& 0.0& 1.0\\
         & $\geq 3g$& 0.0& 1.0\\\midrule
         \multicolumn{4}{c}{$\boldsymbol{v_{kick}}$}\\\midrule
         \texttt{\underline{McFACTSnosur}}&  $1g$&  0.10& $\approx 0.0$\\
         &  $2g-ng$&  0.60& $\approx 0.0$\\
         &  $\geq 3g$&  0.63& $\approx 0.0$\\
         \texttt{\underline{McFACTSfilter}}&  $1g$&  0.10& $\approx 0.0$\\
         &$2g-ng$& 0.60& $\approx 0.0$\\
         &  $\geq 3g$&  0.59& $\approx 0.0$\\
         \texttt{\underline{McFACTSprec}}&  $1g$&  0.01& 0.71\\
         & $2g-ng$& 0.60& $\approx 0.0$\\
         & $\geq 3g$& 0.59& $\approx 0.0$\\
         \texttt{\underline{McFACTSsur}}& $1g$& 0.0& 1.0\\
         & $2g-ng$& 0.0& 1.0\\
         & $\geq 3g$& 0.0& 1.0\\\bottomrule
    \end{tabular}
    \caption{Generated KS Statistic and p-value for the number distributions, n, of $a_{final}$ and $v_{kick}$.}
    \label{tab:kstest}
\end{table}

\subsubsection{\texttt{McFACTSsur}}

For \texttt{McFACTSsur} in Fig. ~\ref{fig:spin_v_kick}, the spin values increase with merger generation. As noted above, the $1g$ merger (gold) distribution is similar to \texttt{McFACTSnosur}. The $a_{\mathrm{final}}$ values for $2g$ mergers (purple) drastically shift compared to the initial \texttt{McFACTSnosur} values and range $a_{\rm final} \sim [0.7, 0.9]$, centering on $a_{\rm final} \sim 0.85$. The $a_{\mathrm{final}}$ range for the $3g+$ mergers (red) spans $a_{\rm final} \sim [0.8, 0.95]$, with values centering just below $a_{\rm final} \sim 0.9$. The right sub-panel shows the clear separation in $a_{\mathrm{final}}$ between merger populations.

\begin{figure*}
    \centering
    \includegraphics[width=\textwidth]{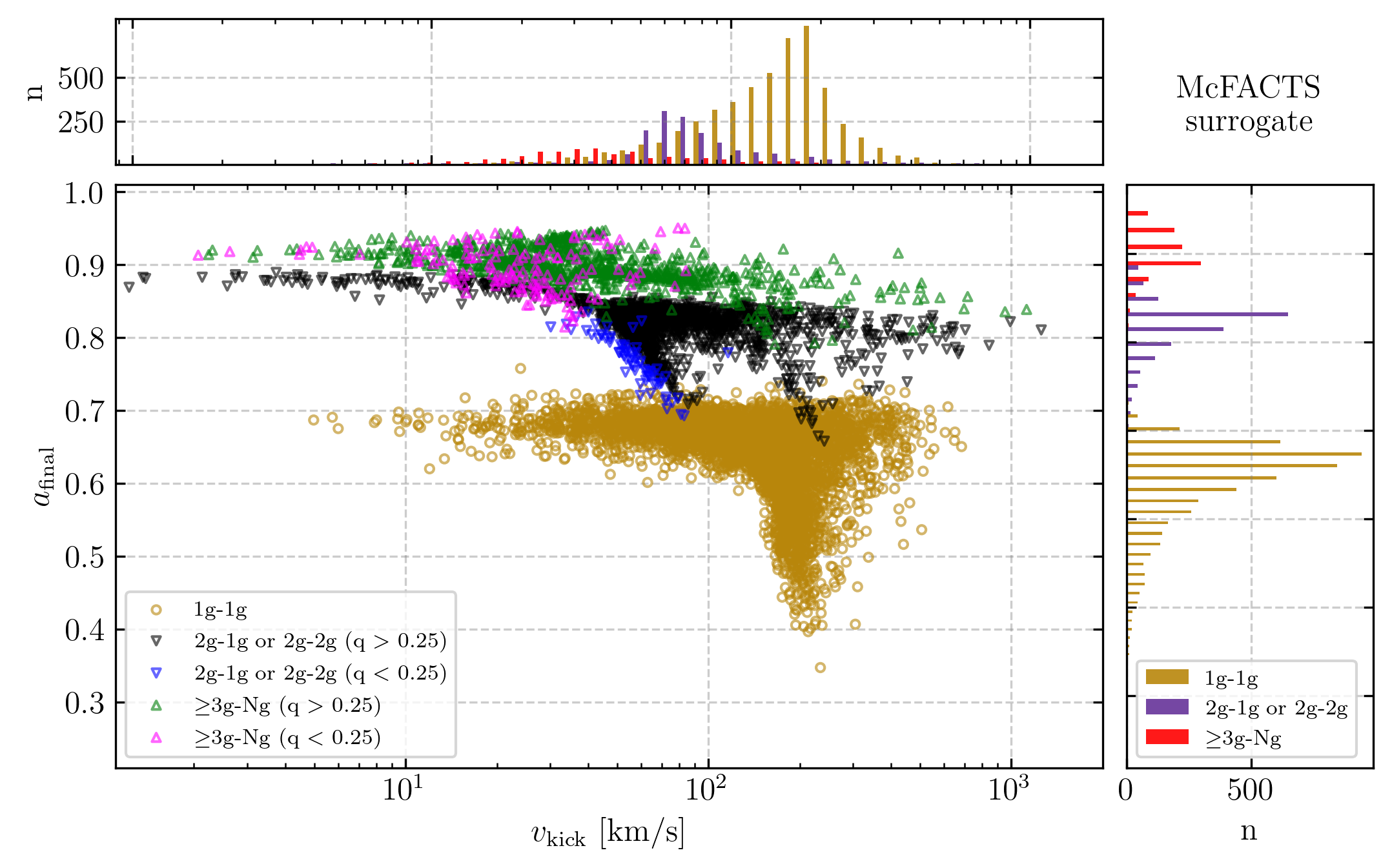}
    \caption{Spin versus kick velocity with each color representing a region above or below the $q=0.25$ cutoff using \texttt{McFACTSsur}. Blue points are $2g-ng$ mergers with $q<0.25$, black points are $2g-ng$ mergers with $q>0.25$, pink points are $3g+$ mergers with $q<0.25$ and green points are $3g+$ mergers with $q>0.25$. Each histogram, $\mathrm{n}$ represents the number density of each axis.}
    \label{fig:spin_cutoff}
\end{figure*}

\subsubsection{\texttt{McFACTS} + Alternatives}

For \texttt{McFACTSprec} in Fig. ~\ref{fig:spin_v_kick}, the $a_{\mathrm{final}}$ distribution features are similar to those found in \texttt{McFACTSsur}. This analytical model was implemented as a computationally efficient way for BBH merger populations to be calculated and to generate distributions of $a_{\mathrm{final}}$ most similar to the extremely accurate surrogate model calculations. Each generation forms separate distributions and remnant $a_{\mathrm{final}}$ increases as generation increases. 
From Table \ref{tab:kstest}, the p-values of $a_{\mathrm{final}}$ for each merger generation are greater than the commonly used significance level, 0.05. This tells us all $a_{\mathrm{final}}$ values calculated for \texttt{McFACTSprec} are statistically significant in matching those calculated for \texttt{McFACTSsur}, indicating they can be used in further simulations investigating hierarchical merger spin populations.

For \texttt{McFACTSfilter} in Fig.~\ref{fig:spin_v_kick}, the merger populations follow the trend of increasing $a_{\mathrm{final}}$ as the generation increases. A random process is applied when resetting the $a_{\mathrm{final}}$ values, where the goal is to assign the correct spin distribution in the most computationally efficient method possible, as opposed to any effort at calculating physical accuracy. While this approach produces $a_{\mathrm{final}}$ values similar \texttt{McFACTSsur}, our applied KS Test shows the distributions cannot be equated to each other. As shown in Table \ref{tab:kstest}, the p-values of $a_{\mathrm{final}}$ for each merger generation are less than the commonly used significance level, 0.05. This tells us none of the $a_{\mathrm{final}}$ values calculated for \texttt{McFACTSfilter} match those calculated for \texttt{McFACTSsur}, indicating they are not accurate enough to be used in further simulations investigating hierarchical merger spin populations.

\begin{figure*}
    \centering
    \includegraphics[width=\textwidth]{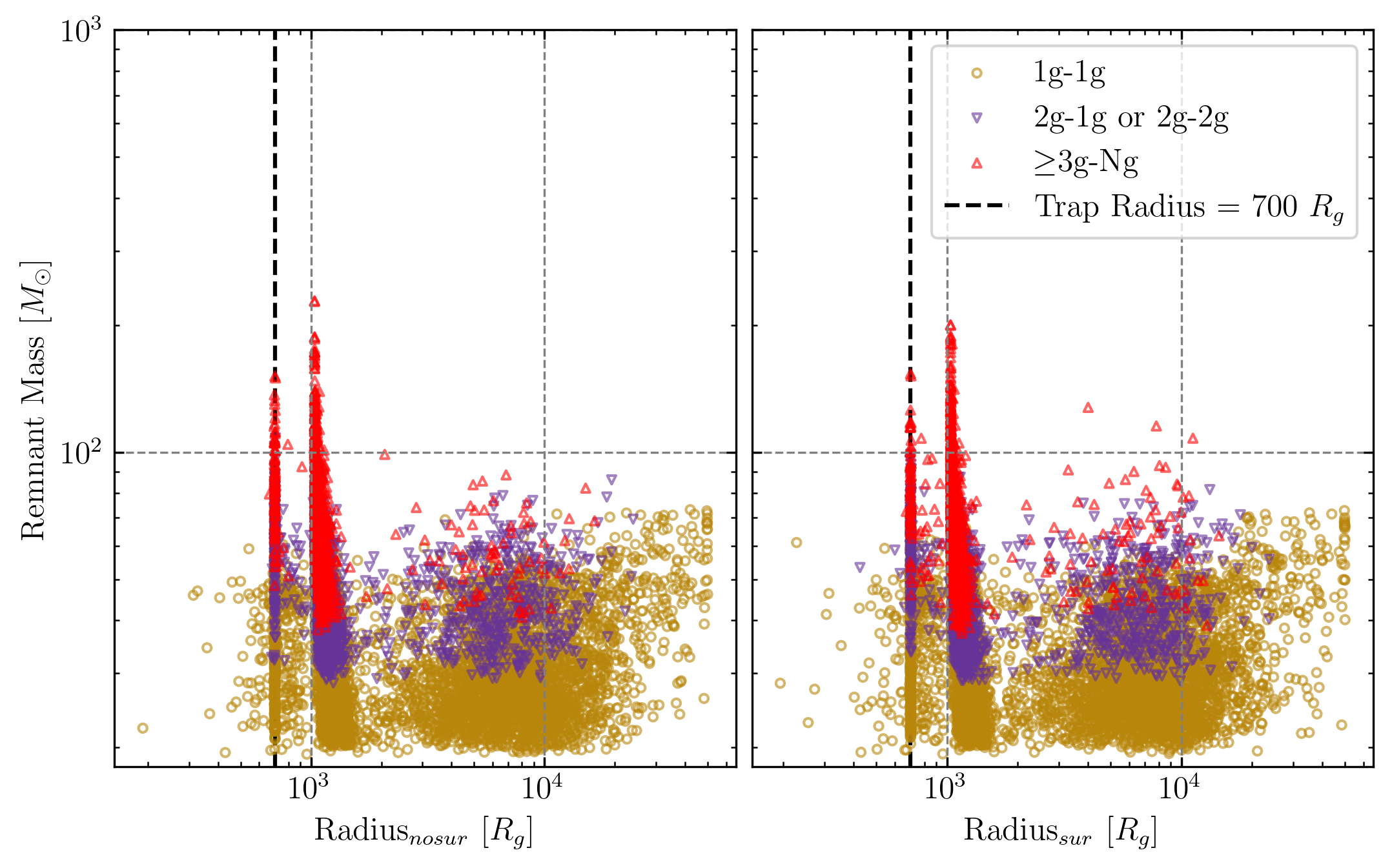}
    \caption{BBH merger remnant masses as a function of radius throughout the disk in units of gravitational radii $R_g$. The left panel is calculated by \texttt{McFACTSnosur}. The right panel is calculated by \texttt{McFACTSsur} after reaching 1,000$M$ in units of binary total mass $M$.}
    \label{fig:mass_radius}
\end{figure*}

\subsection{Kick Velocity}
\label{sec:vkick}

The \texttt{McFACTSnosur} $1g$ $v_{\mathrm{kick}}$ values in Fig. \ref{fig:spin_v_kick} show a broad distribution spanning $v_{\rm kick} \sim [10, 10^3]$km/s with a peak centered just below $v_{\rm kick} \sim 200$km/s. The $2g$ and $3g+$ mergers have a similar broad distribution, but are much more concentrated with their $v_{\mathrm{kick}}$ values centered at $v_{\rm kick} \sim 100$km/s.
Higher generation mergers show similar features such as the wide spread of $v_{\mathrm{kick}}$ values spanning $v_{\rm kick} \sim [10^0,10^3]$km/s and the bimodal peaks made in the $2g$ merger population around $v_{\rm kick} \sim 100$km/s. As shown in Table \ref{tab:kstest}, the p-values of $v_{\mathrm{kick}}$ for each merger generation are $p<0.05$. So $v_{\mathrm{kick}}$ values calculated for \texttt{McFACTSnosur} are not  statistically significant in matching the distribution calculated for  \texttt{McFACTSsur}, indicating their unreliability in accurately investigating the kick velocities of hierarchical BH populations.

The \texttt{McFACTSsur} $1g$ $v_{\mathrm{kick}}$ values in Fig. \ref{fig:spin_v_kick} show a broad distribution ranging $v_{\rm kick} \sim [10, 10^3]$km/s peaking at $v_{\rm kick} \sim 200km/s$. The $2g$ mergers have a broader distribution ranging $v_{\rm kick} \sim [1, 10^3]$ km/s with a peak centered around $v_{\rm kick} \sim 60km/s$. The distribution of the $2g$ mergers creates a bimodal distribution in its $v_{\mathrm{kick}}$ values. The origin of this feature is highlighted in Fig. \ref{fig:spin_cutoff}. The $3g+$ mergers show broad $v_{\mathrm{kick}}$ values ranging $v_{\rm kick} \sim [1,10^3]$km/s with a peak centering around $v_{\rm kick} \sim 30$km/s. These distributions show a clear decrease in $v_{\mathrm{kick}}$ values as the generation increases. 

The \texttt{McFACTSprec} $1g$ $v_{\mathrm{kick}}$ values in Fig. \ref{fig:spin_v_kick} show a broad distribution spanning $v_{\rm kick} \sim [10, 10^3]$km/s with a peak centered just below $v_{\rm kick} \sim 200$km/s. The $2g$ and $3g+$ mergers have a similar $v_{\mathrm{kick}}$ distribution to \texttt{McFACTSsur} but instead have ranges of $v_{\rm kick} \sim [10, 500]$ km/s with peak values centered at $v_{\rm kick} \sim 100$km/s. As shown in Table \ref{tab:kstest}, the p-values of $v_{\mathrm{kick}}$ for $2g$ and $3g+$ merger generations are $p<0.05$. This tells us these two population of $v_{\mathrm{kick}}$ values calculated for \texttt{McFACTSprec} are not statistically significant in matching those calculated for \texttt{McFACTSsur}. The $1g$ mergers, however, have p-values $p>0.05$, indicating their distributions are match those of the \texttt{McFACTSsur} $1g$ mergers with statistical significance. 

The \texttt{McFACTSfilter} $v_{\mathrm{kick}}$ values in Fig. \ref{fig:spin_v_kick} show distributions similar to \texttt{McFACTSprec}. Despite this similarity, Table \ref{tab:kstest} shows the p-values of $v_{\mathrm{kick}}$ for each merger generation are  $p<0.05$. So, none of the $v_{\mathrm{kick}}$ values calculated for \texttt{McFACTSfilter} are statistically significant in matching those calculated for \texttt{McFACTSsur}, indicating unreliability in investigating the kick velocities of hierarchical BH populations.

A prominent feature of \texttt{McFACTSsur} in Fig. \ref{fig:spin_v_kick} is the bimodal distribution of the kick velocities in the $2g$ mergers which arises due to the mass ratios of the BBHs. This figure uses a unique color scheme to highlight specific features found when using the NR surrogate. Blue points are $2g$ mergers with $q<0.25$ centering just below 100km/s, black points are $2g$ mergers with $q>0.25$ taking up most of the $2g$ distribution shape. Pink points are $3g+$ mergers with $q<0.25$ centering just above 10km/s, and green points are $3g+$ mergers with $q>0.25$ taking up most of the $3g+$ distribution shape.

\begin{figure*}
    \centering
    \includegraphics[width=\textwidth]{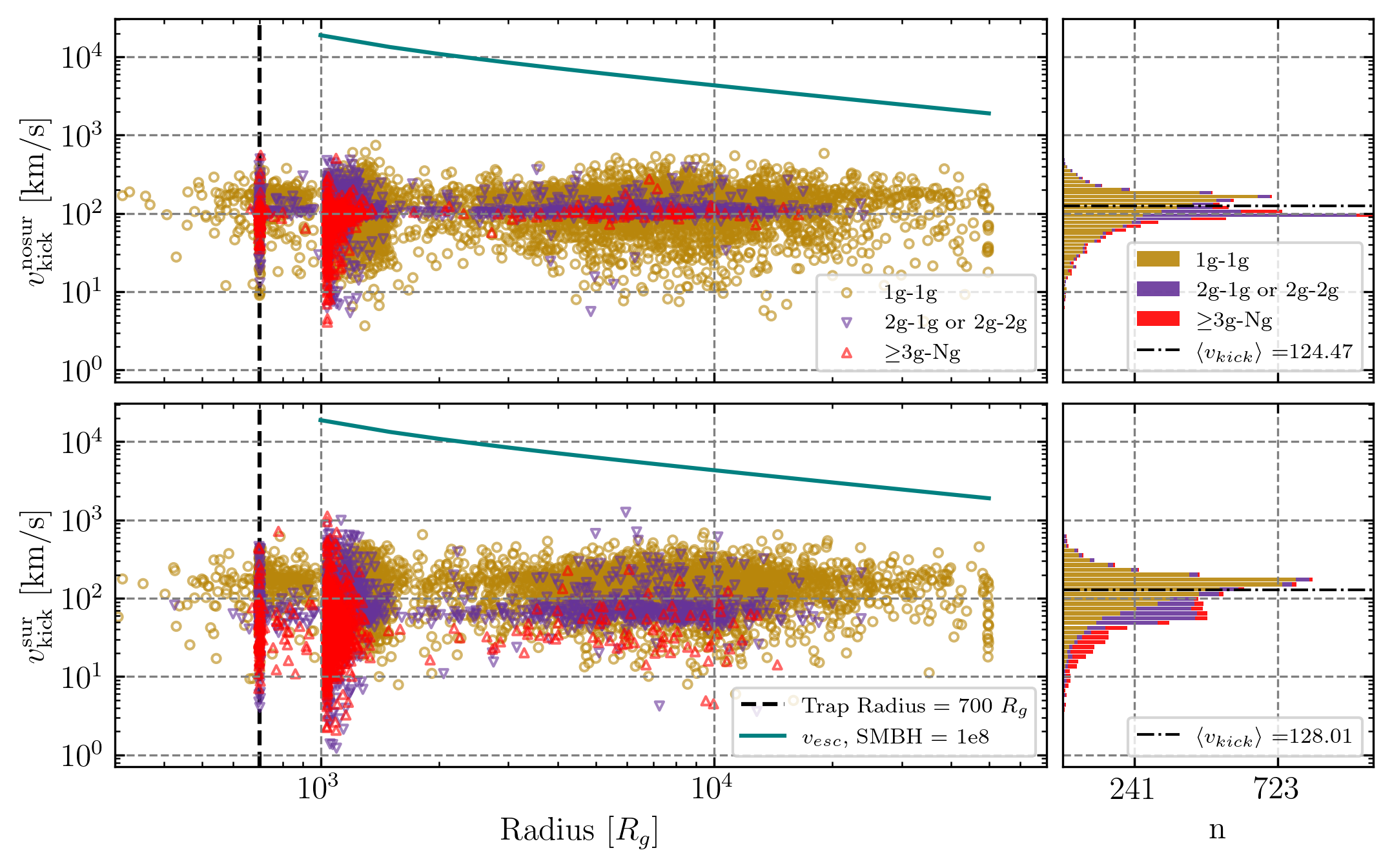}
    \caption{The left panels show the BBH merger remnant kick velocity, $v_{\mathrm{kick}}$, as a function of disk radius, $R_g$. The top left panel shows the results from the default version of McFACTS. The bottom left panel corresponds to a default run with the addition of the NR surrogate model after the BBH reaches a binary separation of 1000$M$ in units of total binary mass. The right panels indicate the number of merger remnants as a function of remnant $v_{\mathrm{kick}}$. The teal line indicates the escape velocity as a function of radius.}
    \label{fig:kick_radius}
\end{figure*}

The top row of Fig.~\ref{fig:kick_radius} shows \texttt{McFACTSnosur} and the bottom row shows \texttt{McFACTSsur} with the left panel showing $v_{\mathrm{kick}}$ as a function of disk radius with three distinct populations similar to those seen in Fig.~\ref{fig:mass_radius}. The right panel shows the distribution of $v_{\mathrm{kick}}$ for the merger remnants. The number distribution of \texttt{McFACTSnosur} in Fig.~\ref{fig:kick_radius} shows an average kick of $\langle{v_{\mathrm{kick}} \rangle}$=127.17 $km/s $ with many of the higher generation mergers having a $v_{\mathrm{kick}}$ close to the average. The number distribution of \texttt{McFACTSsur} shows a decrease in average kick $\langle{v_{\mathrm{kick}}}\rangle$=119.21 $km/s$, however when applying the surrogate model, the higher generation mergers tend to have a lower $v_{\mathrm{kick}}$. In addition to this decrease in kick velocities by the higher generation remnants, we can see there is a group of $3g+$ mergers with $v_{\mathrm{kick}} \approx$ $10^1 km/s$, showing a drastic decrease in the overall population $v_{\mathrm{kick}}$. 

In each model, the teal line represents the escape velocity, $v_{esc}$ as a function of radius across the disk. We can see with \texttt{McFACTSnosur}, all of the merger remnants generate a $v_{\mathrm{kick}}$ lower than the $v_{esc}$ meaning remnants are always kept in the nucleus, allowing for high efficiency of hierarchical mergers. With the addition of NR surrogate, we can see the $v_{\mathrm{kick}}$ values decrease significantly in higher generation remnants, allowing them to remain in the nucleus with the possibility of recapture within the disk lifetime (or in a future AGN episode). 

\subsection{Mass}

Figure ~\ref{fig:mass_radius} shows the BBH merger remnant masses versus the radius of the disk, in units of gravitational radii $R_g$, which group into three distinct population clumps for the higher generation mergers. We can see, going left to right in the first plot, the first population gathers along the migration trap, a region in which the disk torques limit migration, set at $700R_g$, the second population gathers along the migration swamp, a region in which the disk density rapidly changes and slows migration, forming slightly above $10^3R_g$, and the third population gathers across a majority of the disk at ~$[10^3,10^4]R_g$. These three regions are the locations of BBH merger remnants during the AGN's life cycle. Comparing \texttt{McFACTSnosur}, the left plot, and \texttt{McFACTSsur}, the right plot, we can see the population distribution has a few edge cases in which the merger masses are slightly higher and further into the disk, but the masses across the disk generally remain the same. Thus, updating our merger treatment has only small effects on the masses and locations of mergers within the AGN disk telling us the change in spin does not effect BBH disk location.

\subsection{Spin 2}

\begin{figure*}
    \centering
    \includegraphics[width=\textwidth]{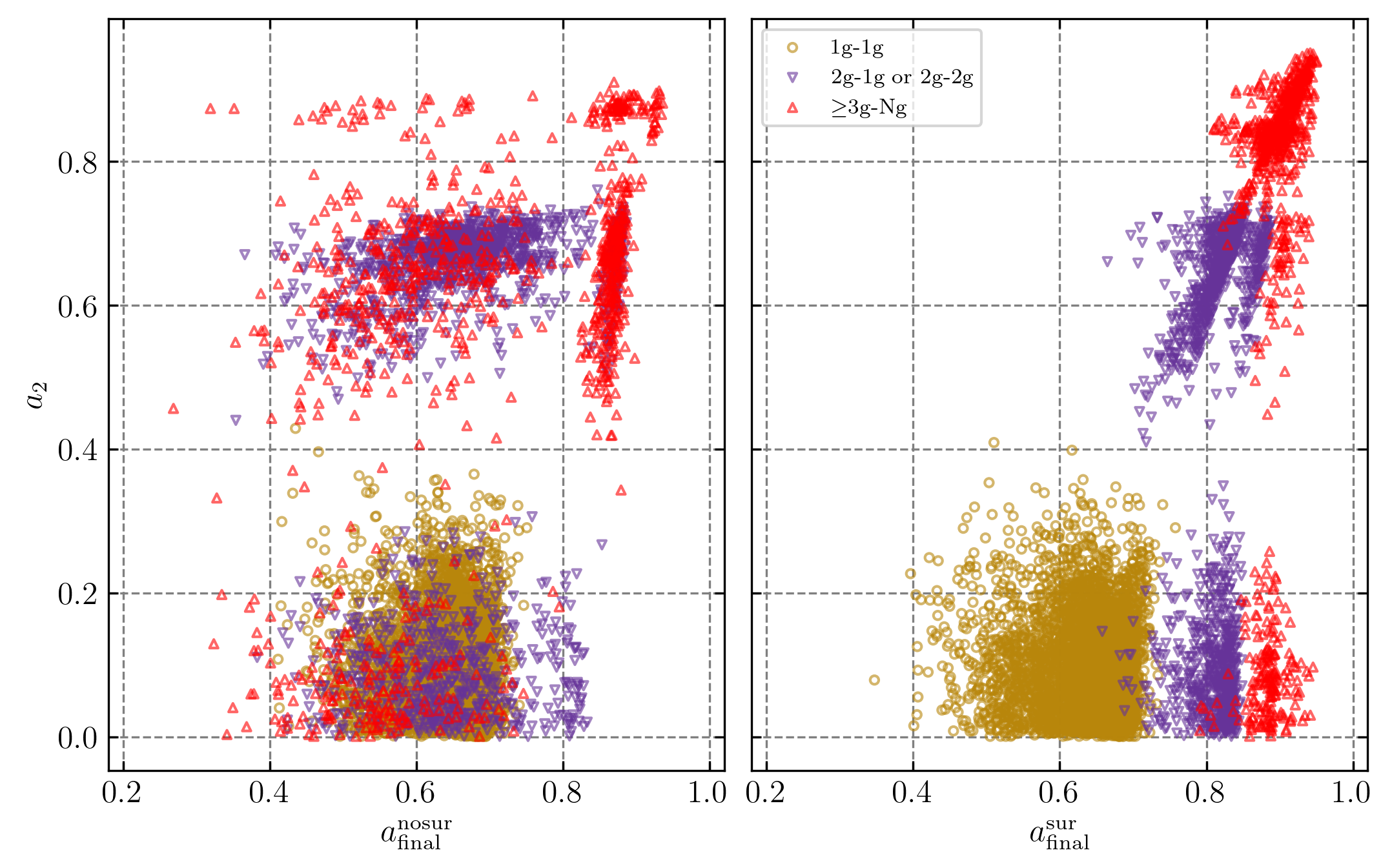}
    \caption{The spins of the secondary object in a BBH merger, $a_{2}$, as a function of the final remnant spin, $a_{\mathrm{final}}$.}
    \label{fig:a2_afinal}
\end{figure*}

In Fig. \ref{fig:a2_afinal}, we plot $a_2$ as a function of $a_{\mathrm{final}}$. Lower mass BHs are disproportionately likely to be $1g$, so we can use $a_2$ to better understand why \texttt{McFACTSnosur} calculations are so poor when predicting final spins compared to \texttt{McFACTSsur} since \texttt{McFACTSnosur} seems to produce results that do not significantly differ from \texttt{McFACTSsur} on $1g-1g$ mergers. 

When comparing \texttt{McFACTSsur} and \texttt{McFACTSnosur}, both methods show higher generation mergers tend to increase in spin value as the generation increases. However, the diverging point of these plots can be seen in the high spins of the high generation mergers. \texttt{McFACTSsur} tends to create a clear trend between the high generation mergers while \texttt{McFACTSnosur} tends to have a more randomized range of values spread across our spin values. The $1g$ remnants tend to gather as one group with $a_2 < 0.4$ whereas the higher generation remnants tend to separate into two distinct groups, high $(a_2> 0.4)$ and low $(a_2< 0.4)$ spin. This separation between high and low spins for the high generation mergers is the biggest change in the spin values from the surrogate implementation.

While spin of the merger remnants increases with higher generation, regardless of method, the surrogate finds consistently higher spins for $1g-ng$ mergers, compared with \texttt{McFACTSnosur}; in addition, the $2g-2g$ and $3g-2g$ mergers produce modestly larger spin remnants using \texttt{McFACTSsur} compared to \texttt{McFACTSnosur}, while the $3g-3g$ remnants are the only ones with consistently high spin in both models. Therefore, we suspect the underlying differences are due to both mass ratio and progenitor spin.

\subsection{Effective Spin}

GW observations suggests that the effective spin distribution broadens as $q$ decreases among BBH merger populations (\cite{gwtc5pop2026}). This relationship has been investigated in other GW channels but isolated binaries in particular have proved challenging to reproduce this broadening effect (\cite{ban_ole2026}, \cite{b_kal2024}).  Initial results from LVK O3 suggested that there was an anti-correlation in $[q, \chi_{\rm{eff}}]$  \citep{callister2021} but more recently, with O4, the $\chi_{\rm eff}$ distribution broadens (to both positive and negative values) as the mass ratio ($q$) decreases  \citep{gwtc5pop2026}. To reflect this change we have changed parameter \texttt{frac$\_$bin$\_$retro} from 0.0 to 0.5 in Table~\ref{tab:default}, indicating that half of all BBH will form retrograde.

\begin{figure*}
    \centering
    \includegraphics[width=\textwidth]{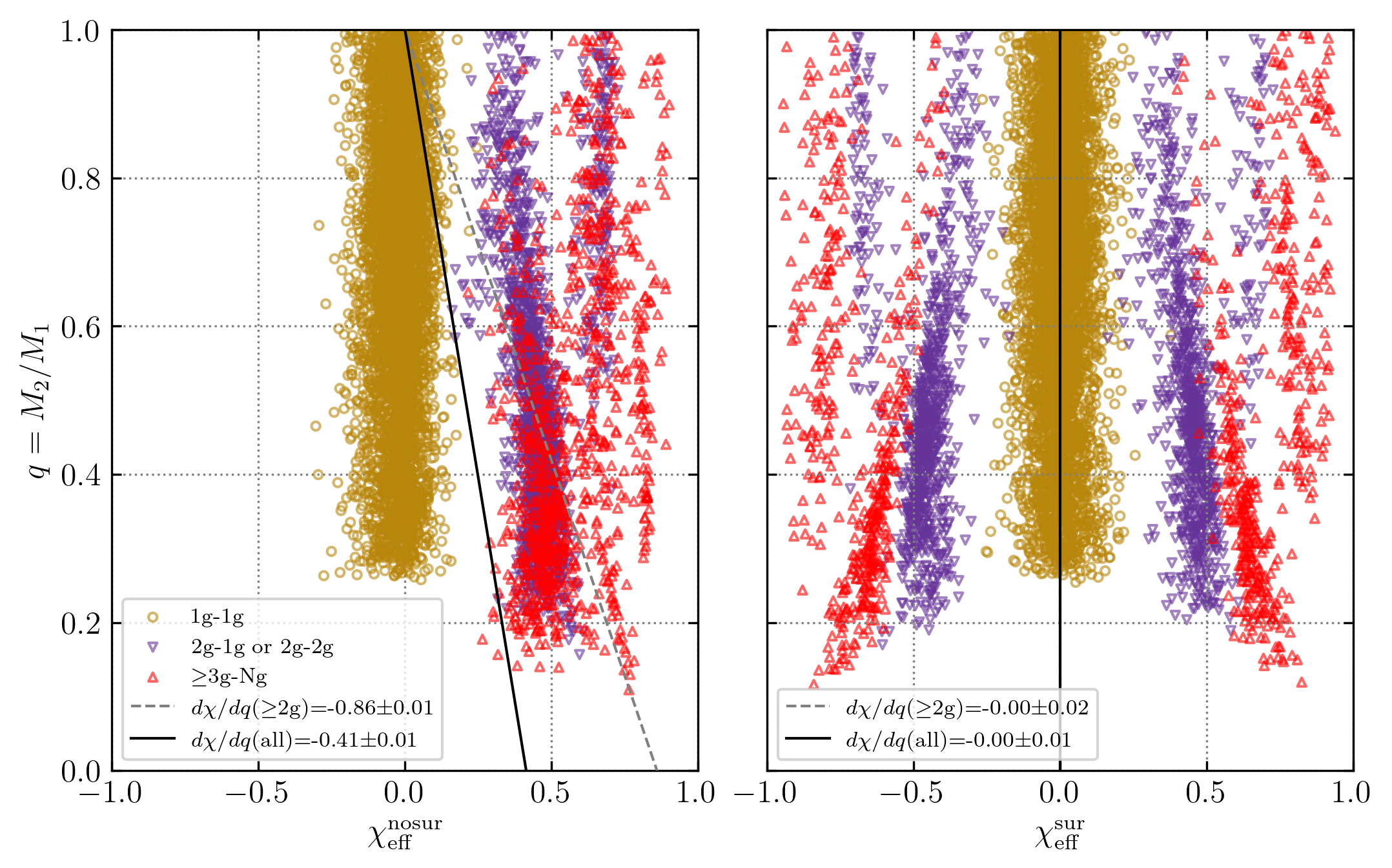}
    \caption{Distribution of mass ratio, $q$, for BBH mergers as a function of $\chi_{\rm eff}$.}
    \label{fig:q_chieff}
\end{figure*}

Figure ~\ref{fig:q_chieff} plots the the BBH mass ratio $q$, as a function of its effective spin, $\chi_{\rm eff}$. On the LHS we show the $[q,\chi_{\rm eff}]$ plot using the old \texttt{frac$\_$bin$\_$retro}$=0.0$ as in \citep{McFACTS_II} for \texttt{McFACTSnosur}. On the RHS we show \texttt{frac$\_$bin$\_$retro}$=0.5$ and \texttt{McFACTSsur}. In both cases, the $1g$ mergers (gold) form a population distribution centered around $\chi_{\rm eff} \sim 0$ spanning $\chi_{\rm eff} \sim [-0.25, 0.25]$ with $q$ $> 0.25$ resulting from our default choices of IMF and spins. The $2g$ mergers form two positive distributions and two negative distributions spanning across two different ranges. Both positive and negative distributions of the mass ratios are $q > 0.1$ and $q > 0.4$; these distributions center around $\chi_{\rm eff}=\pm 0.4$ and $\chi_{\rm eff}=\pm 0.6$. The $3g+$ mergers form similar overlapping population clumps in \texttt{McFACTSnosur}, but cleanly separate at slightly larger $\chi_{\rm{eff}}$ with \texttt{McFACTSsur}. In \texttt{McFACTSsur}, the $q$ values for the $3g+$ mergers decrease and reach nearly $q > 0.1$ and $q > 0.2$, respectively. This decrease in $q$ follows the increase in mass as BBH generation increases since larger masses will generate lower mass ratios with smaller, early generation, objects. The $\chi_{\rm{eff}}$ value of $3g+$ mergers increases to the point of no longer overlapping with the $2g$ mergers and ranges from $[0.5,0.95]$. 

With the addition of \texttt{McFACTSsur}, we see the spins, $\chi_{\rm eff}$, and kick velocities change significantly, while the masses, disk location, and escape fractions do not change significantly. Improving the details of BBH merger calculations within AGN disks shows that this hierarchical merger channel is remarkably efficient. 


\section{discussion}

We find that it is possible to considerably improve on the accuracy of final mass, spin, and kick velocity measurements over default analytic models (\texttt{McFACTSnosur}). 

In the case of studies of hierarchical mergers in other channels,  particularly ones where remnant kick velocities are important, we recommend using alternative methods to analytic models such as \cite{TichyM2008} because of its inaccuracies, particularly for higher generation mergers. In order to calculate physically accurate spins and kick velocities for higher generation remnants in the AGN channel, \texttt{McFACTSsur} should be considered. In cases where only spins are necessary, we recommend \texttt{McFACTSprec} for its computational efficiency compared to \texttt{McFACTSsur}. Despite its computational expense, \texttt{McFACTSsur} is recommended for cases of high mass high spin calculations of remnants in AGN disks. 

\subsection{BH populations in AGN}

Implementing \texttt{McFACTSsur} has shown that including numerical relativity when calculating BBH merger remnant parameters is essential in accurately determining BBH merger properties and possible population dynamics within AGN disks. Parameters such as kick velocity and $\chi_{\rm{eff}}$ are directly effected by BBH merger remnant mass and spin, so any discrepancies in remnant mass and spin calculations can compound as BBH generation increases. Our approach has also shown while \texttt{McFACTSsur} is computationally efficient as a surrogate to NR  calculations, it is still a computationally expensive tool when applied to large scale population analyses like \texttt{McFACTS}. So, in order to save on computational efficiency while generating results  comparable to \texttt{McFACTSsur}, we implemented \texttt{McFACTSprec}, which was specifically designed to be a computationally efficient method of calculating remnant parameters on large-scale population analysis codes. To further allow for computational efficiency, we also implemented a hard coded filter, \texttt{McFACTSfilter} to detect and adjust spin calculations to better match the distribution of results found using \texttt{McFACTSsur} spin calculations; the \texttt{McFACTSfilter} option is intended solely for users willing to sacrifice physical accuracy for the highest computational efficiency possible.

The increase in accuracy of \texttt{McFACTSsur} spin calculations show a decrease in  $v_{kick}$ as BHs increase in generation. If higher generation BHs do not leave the disk over the AGN lifetime, then sufficiently long-lived AGN serve as an excellent environment for intermediate mass black hole (IMBH) growth and potential extreme mass ratio inspiral (EMRI) detections. The increase in accuracy of \texttt{McFACTSsur} spin calculations has also improved the accuracy of $\chi_{\rm eff}$ calculations, showing higher generation hierarchical mergers within AGN disks are more likely to align and anti-align with the orbital angular momentum of the disk. 

\subsection{LVK Detections}

\begin{figure*}
    \centering
    \includegraphics[width=\textwidth]{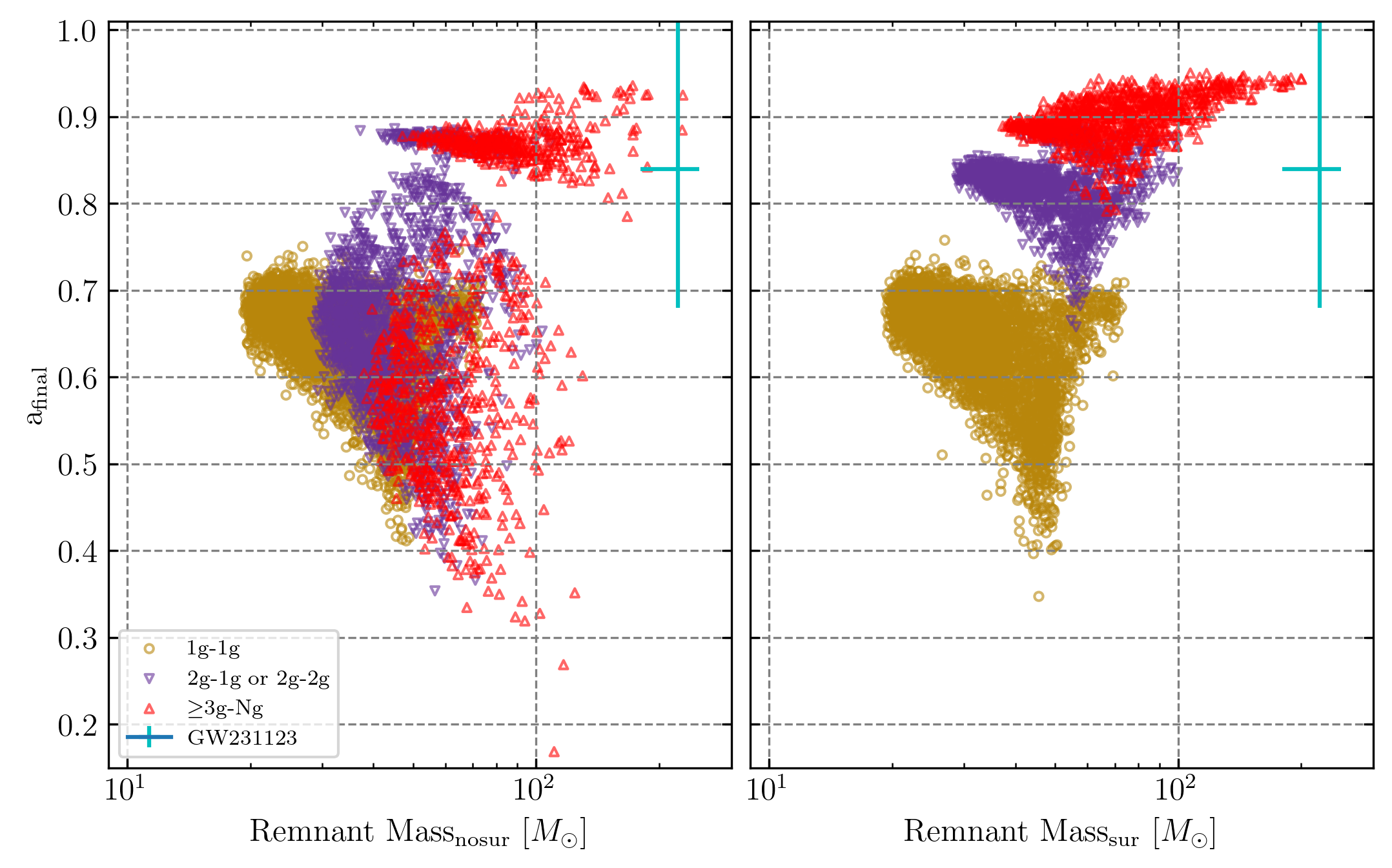}
    \caption{Remnant spin versus remnant mass of \texttt{McFACTSnosur}, on the left, and \texttt{McFACTSsur}, on the right. Also depicted is the BBH merger GW231123 highlighted in cyan.}
    \label{fig:231123}
\end{figure*}

Improving mass and spin calculations with \texttt{McFACTSsur} has shown BBH merger populations similar to those found in LVK O4 with high mass and spin. The largest total mass merger detected by LVK so far is GW231123. In \citep{gw231123}, by combining results from five equally weighted and mixed models with median values of the 90\% highest posterior density (HPD) credible interval, the following parameters result: final mass of $M_f = 222_{-42}^{+28} M_\odot$, a final spin of $\chi_f = 0.84_{-0.16}^{+0.08}$, progenitor masses of $m_1 = 137^{+23}_{-18} M_\odot$ and $m_2 = 101^{+22}_{-50} M_\odot$ with $\chi_{eff} =0.32^{+0.25}_{-0.41}$. These masses put the progenitors of GW231123 well within the upper mass-gap and are thought to be the result of a chain of hierarchical merger events. 

Figure \ref{fig:231123} shows $a_{\rm final}$ versus remnant mass for our default run, with the final mass and spin of GW231123 from \cite{gw231123}. Evidently, while we did not attempt to fit the parameters of GW231123 with our generic default run, we are close enough. The parameters found for GW231123 seem  difficult to match, at high efficiency, in formation channels other than AGN  (\cite{delfavero_prospects_2025}).

\subsection{Implications for GW Channels}

Observations of BBH spins are excellent tests of the AGN channel. An understanding of the BBH population spins can give more accurate predictions on the BBH merger rates in AGN, AGN  lifetimes, and AGN disk size (\cite{mck2018}, \cite{ford_using_2025}). Implementing  \texttt{McFACTSsur} shows that AGN lifetimes $\sim 1\rm{Myr}$ efficiently produce high generation mergers with remnants that have high spins consistent with GW231123. The high spins that persist throughout the AGN timescale, show that it is possible to generate conditions for hierarchical mergers during relatively short lived AGN disks. Generating events such as these reinforces the idea that some AGN disks are geometrically thin and gas damping can be very efficient  disks (\cite{mckFord24}). 

The kick velocities of BBH populations help us determine if remnants can leave the AGN disk, or the nucleus itself, effectively determining the efficiency of the mass merger hierarchy. We find a decrease in kick velocities among higher generation mergers, implying that remnants are unlikely to be kicked far from the disk mid-plane. This suggests that the mass merger hierarchy is very efficient, with a higher density of objects retained by and migrating in the disk over its lifetime. The buildup of high mass objects at migration traps and swamps indicates that these regions are excellent environments for IMBH growth and possible EMRI detections. 

Finally, analytical calculations for higher generation BBH mergers should be used cautiously when determining the spins and kick velocities of BBH populations, for any hierarchical merger environment (particularly those with low escape velocities, but this applies to all channels). Studies that focus on hierarchical mergers in LVK detections should fully consider using a numerical approach when calculating remnant kick velocities due to the low escape velocities found in globular clusters. For example, our methods show that the decrease in kick velocities of higher generation mergers could potentially increase the number of objects found in globular clusters. While analytical models are sufficient in generating approximate results, in the context of BBH merger populations and their origins, numerical methods prove to excel in calculating parameters in which analytical models fall short.


\section{acknowledgments}

S.R. is supported by the CUNY GC MS Astrophysics program and thanks the Flatiron Institute for hospitality and Guest Researcher access during the completion of this research. The Flatiron Institute is a division of the Simons Foundation. K.M. is supported by NASA through the NASA Hubble Fellowship grant \#HST-HF2-51562.001-A awarded by the Space Telescope Science Institute, which is operated by the Association of Universities for Research in Astronomy, Incorporated, under NASA contract NAS5-26555. V.D. acknowledges support from the Natural Sciences and Engineering Research Council of Canada (NSERC) under grant RGPIN-2023-05511 and the Research Corporation for Science Advancement’s Scialog Initiative under grant No. SALSST-2024-089b with support from the Brinson Foundation. EM would like to thank the CUNY GC MS Astrophysics program and the CCA GW group for support and useful comments. EM thanks the CCA at the Flatiron Institute for hospitality while a portion of this research was carried out. The Flatiron Institute is a division of the Simons Foundation. BM, KESF and HEC are supported by NSF AST-2206096. BM \& KESF are supported by NSF AST-1831415 and Simons Foundation Grant 533845. The Flatiron Institute is supported by the Simons Foundation. 

\textit{Software:} This work made use of the following software packages: \texttt{astropy} \citep{astropy:2013,astropy:2018,astropy:2022,astropy_14202906}, \texttt{matplotlib} \citep{Hunter:2007}, \texttt{numpy} \citep{numpy}, \texttt{pandas} \citep{mckinney-proc-scipy-2010,pandas_13819579}, \texttt{precession} \citep{gerosa_kesden2016}, \texttt{python} \citep{python}, \texttt{scipy} \citep{2020SciPy-NMeth,scipy_14880408}, \texttt{McFACTS} \citep{McFACTS_I,McFACTS_II,McFACTS_III,McFACTS_IV}, and \texttt{scikit-learn} \citep{scikit-learn,sklearn_api,scikit-learn_15606905}.

This research has made use of the Astrophysics Data System, funded by NASA under Cooperative Agreement 80NSSC21M00561.

Software citation information aggregated using \texttt{\href{https://www.tomwagg.com/software-citation-station/}{The Software Citation Station}} \citep{software-citation-station-paper,software-citation-station-zenodo}.

\textit{Data Availability:} \texttt{McFACTS} is available at \href{https://github.com/mcfacts}{https://github.com/mcfacts}, and
this work was completed with the version at git hash \texttt{dcce980}. The data used in this paper is available at Zenodo, DOI:\href{10.5281/zenodo.23383305}{10.5281/zenodo.23383305} (\cite{sray_zenodo2026}). This work was also completed with a currently proprietary NR surrogate, which will be publicly available in the future and be detailed in an upcoming publication.

\bibliography{bibliography}{}
\bibliographystyle{aasjournalv7}

\end{document}